\documentclass[astrosymb]{aastex631}

\hypersetup{linkcolor=red,citecolor=green,filecolor=cyan,urlcolor=magenta}
\newcommand\Msun{M_{\odot}}

\submitjournal{ApJL}

\graphicspath{{./}{figures/}}

\begin{document}

\title{Expanding the time domain of multiple populations: evidences of nitrogen variations in the $\sim1.5$~Gyr~old star cluster NGC~1783}

\correspondingauthor{Mario Cadelano}
\email{mario.cadelano@unibo.it}

\author[0000-0002-5038-3914]{Mario Cadelano}
\affil{Dipartimento di Fisica e Astronomia “Augusto Righi'', Universit\`a degli Studi di Bologna, Via Gobetti 93/2, 40129 Bologna, Italy}
\affil{INAF-Osservatorio di Astrofisica e Scienze dello Spazio di Bologna, Via Gobetti 93/3 I-40129 Bologna, Italy}

\author[0000-0003-4237-4601]{Emanuele Dalessandro}
\affiliation{INAF-Osservatorio di Astrofisica e Scienze dello Spazio di Bologna, Via Gobetti 93/3 I-40129 Bologna, Italy}

\author[0000-0002-2744-1928]{Maurizio Salaris}
\affiliation{Astrophysics Research Institute, Liverpool John Moores University, 146 Brownlow Hill, Liverpool L3 5RF,UK}
\affiliation{INAF - Osservatorio Astronomico d'Abruzzo, via M. Maggini, 64100, Teramo, Italy}

\author[0000-0001-5679-4215]{Nate Bastian}
\affiliation{Astrophysics Research Institute, Liverpool John Moores University, 146 Brownlow Hill, Liverpool L3 5RF,UK}
\affil{Donostia International Physics Center (DIPC), Paseo Manuel de Lardizabal, 4, E-20018 Donostia-San Sebasti\'an, Guipuzkoa, Spain}
\affil{IKERBASQUE, Basque Foundation for Science, 48013, Bilbao, Spain}

\author[0000-0001-9158-8580]{Alessio Mucciarelli}
\affil{Dipartimento di Fisica e Astronomia “Augusto Righi'', Universit\`a degli Studi di Bologna, Via Gobetti 93/2, 40129 Bologna, Italy}
\affil{INAF-Osservatorio di Astrofisica e Scienze dello Spazio di Bologna, Via Gobetti 93/3 I-40129 Bologna, Italy}

\author[0000-0003-4746-6003]{Sara Saracino}
\affiliation{Astrophysics Research Institute, Liverpool John Moores University, 146 Brownlow Hill, Liverpool L3 5RF,UK}

\author[0000-0001-7110-6775]{Silvia Martocchia}
\affiliation{Department of Astrophysics/IMAPP, Radboud University, P.O. Box 9010, 6500 GL Nijmegen, The Netherlands}

\author[0000-0001-9478-5731]{Ivan Cabrera-Ziri}
\affiliation{Astronomisches Rechen-Institut, Zentrum f\"ur Astronomie der Universit\"at Heidelberg, M\"onchhofstra{\ss}e 12-14, D-69120 Heidelberg, Germany}

\begin{abstract}
We present the result of a detailed analysis of HST UV and optical deep images of the massive and young ($\sim1.5$~Gyr) stellar cluster NGC~1783 in the Large Magellanic Cloud. This system does not show evidences of multiple populations (MPs) along the red giant branch (RGB) stars. However, we find that the cluster main-sequence (MS) shows evidence of a significant broadening ($50\%$ larger than what is expected from photometric errors) along with hints of possible bimodality in the MP sensitive ($m_{F343N}-m_{F438W},m_{F438W}$) color-magnitude-diagram (CMD). Such an effect is observed in all color combinations including the $m_{F343N}$ filter, while it is not found in the optical CMDs. 
This observational evidence suggests we might have found light-element chemical abundance variations along the MS of NGC~1783, which represents the first detection of MPs in a system younger than 2 Gyr.  
A comparison with isochrones including MP-like abundances shows that the observed broadening is compatible with a N abundance enhancement of $\Delta([N/Fe]) \sim 0.3$. 
Our analysis also confirms previous results about the lack of MPs along the cluster RGB. However, we find that the apparent disagreement between the results found on the MS and the RGB is compatible with the mixing effects linked to the first dredge-up. This study provides new key information about the MP phenomenon and suggests that star clusters form in a similar way at any cosmic age.

\end{abstract}

\section{Introduction}
\label{sec:intro}
Globular clusters (GCs) exhibit star-to-star variations in their light-element content (e.g., \citealt{carretta09}). In fact, while some GC stars have the same light-element abundances as the field at the same metallicity (first population - FP), others show enhanced N and Na along with depleted C and O abundances (second population - SP). Such anomalies are readily observable also by using color-magnitude diagrams (CMDs) involving specific near-UV filters sensitive to OH, CN and CH molecular bands (e.g. \citealt{sbordone11,piotto15}). 
The manifestation of such light-element inhomogeneities is referred to as multiple populations (MPs).
A number of scenarios have been proposed over the years to explain the formation of MPs (e.g., \citealt{decressin07,dercole08,denissenkov14,gieles18}), however, their origin is still strongly debated (see \citealt{bastian18,gratton19} for a recent review).

The MP phenomenon appears to be ubiquitous. In fact, not only all massive and old Galactic GCs host MPs (e.g., \citealt{piotto15,milone17}), but MPs are also observed in the Large and Small Magellanic Clouds (LMC) old stellar clusters \citep{mucciarelli09,dalessandro16}, in GCs in dwarf galaxies such as Fornax \citep{larsen14} and in the M31 GC systems \citep{schiavon13}, and there are strong indications (though based on integrated quantities) that they are a common property of stellar clusters in massive elliptical galaxies \citep[e.g.][]{chung11}. Conversely, several works based on photometric and spectroscopic analysis of red giant branch stars (RGB -- e.g. \citealt{mucciarelli08,mucciarelli14, 
martocchia18}) suggest that
massive clusters younger than $\sim2$ Gyr do not show 
any inhomogeneity in their light-element content. In fact, NGC~1978 in the LMC is the youngest cluster ($t\sim2$ Gyr) found to host sub-populations with light-element chemical variations so far (\citealt{martocchia18,saracino20muse}).
It is worth stressing that, while young clusters ($<2$ Gyr) do show features in their optical CMDs (e.g., extended main sequence turn-offs, dual main sequences) that are not consistent with the classical notion 
of a simple stellar population, 
these features are not due to abundance variations \citep{mucciarelli08,mucciarelli09} 
{but are likely due to stellar rotation  \citep[][]{bastian09,kamann20,kamann21}}. 
Hence, while they may be in principle related to MPs, the underlying cause is different.

The lack of MPs in young ($<2$ Gyr) clusters is completely unexpected and 
inconsistent with predictions for all theories of MP formation.  
We note also that an age of $2$~Gyr corresponds to a formation redshift of $z=0.17$, well past the peak epoch of GC formation (e.g. \citealt{brodie06}). 
One possible explanation for the lack of MPs in young age ($<2$ Gyr) clusters is that old GCs were simply much more massive at birth than those systems that do not show abundance spreads. 
Such large masses may allow GCs to retain stellar ejecta of stars within them and also to accrete pristine gas from their surroundings. 
Indeed cluster mass is found to play a significant role in shaping the properties of MPs in GCs \citep{carretta10,milone17}.
One alternative explanation is that light-element variations do exist also within young clusters, 
but they are difficult/impossible to observe along the RGB, where they have been typically searched for. 
In fact, \citet{salaris20} have recently shown that the mixing effect associated to
the first dredge-up can have a differential impact on the surface chemical abundances of FP and SP RGB stars and it is able to smooth out their initial N abundance differences with increasing efficiency for decreasing ages. 

To finally establish the presence of MPs in young massive clusters, it is therefore 
key to search for MPs along their MS. 
%We note that in the ancient GCs, clear splits in the MS, due to the MP phenomenon, are a common feature in all systems observed to sufficient depth (e.g., \citealt{piotto07}), down to at least $0.3\Msun$-$0.4\Msun$.
To this aim, we have started a comprehensive study 
of the young ($\sim1.5$~Gyr, \citealt{mucciarelli07,zhang18}) cluster NGC~1783 in the LMC.
This system represents an optimal choice in this context as
it is quite massive ($\sim2\times10^5
M_{\odot}$; \citealt{song21}), it is located in a region of the LMC characterized by low extinction ($A_V<0.1$ mag) and by a negligible field contamination. In addition, previous photometric and spectroscopic studies of RGB stars 
\citep{mucciarelli07,cabreraziri2016young,zhang18,martocchia18,martocchia21} suggest that this cluster does not host MPs.
Here we present the results of the first detailed screening of the cluster main sequence (MS) obtained through deep 
Hubble Space Telescope (HST) optical and UV MP sensitive photometry.

The Letter is structured as follows. In Section 2 the adopted data-set and data-reduction procedures are described. Section 3 reports on the MP analysis in the CMD and a comparison with theoretical models. Finally we discuss the main results in Section 4.

\section{Observations and data analysis}

\subsection{Data-set and data reduction}
\label{sec:datared}
This work is based on observations obtained with the UVIS channel of the Wide Field Camera 3 (WFC3) and the Advanced Camera for Surveys (ACS) aboard HST. 
The main data-set is composed of proprietary WFC3 data obtained under GO 16255 (PI: Dalessandro) and consists of 8 images acquired with the F343N filter ($6\times3086$ s and $2\times3095$ s) and 6 images acquired with the F438W filter ($6\times938$ s). 
%As discussed in \citep{niederhofer17a,niederhofer17b,martocchia19,cabrera-ziri20} the combination of near-UV and optical observations obtained with the F343N and F438W filter provide an efficient photometric tool to distinguish stars having a different N content also along the cluster main sequence.
These data were then combined with archival images obtained under GO 10595 (PI: Goudfrooij) and GO 12557 (PI: Girardi). These complementary data-sets consist of 3 ACS images acquired with each of the following filters, F435W ($2\times340$ s and $1\times90$ s), F555W ($2\times340$ s and $1\times40$ s), F814W filter ($2\times340$ s and $1\times8$ s) and 3 WFC3 images acquired with the F336W filter ($2\times1190$ s and $1\times1200$ s). 

The photometric analysis of the entire data-set was performed by using \texttt{DAOPHOT IV}
\citep{stetson87} and following the approach adopted in previous works \citep[see][]{dalessandro18a,dalessandro18b,cadelano19,cadelano20psr}. 
Briefly, tens of bright and isolated stars have been selected in each frame to model the point spread function (PSF), which has been eventually applied to all sources detected in each image above $3\sigma$, where 
$\sigma$ is the standard deviation of the background counts. We then created a master list composed 
of stars detected in at least half of the deep F343N and F438W images. At the corresponding positions of stars in this final master-list, a fit was forced with \texttt{DAOPHOT/ALLFRAME} \citep{stetson94} in each frame of the two data-sets. For each star thus recovered, multiple magnitude estimates obtained in each chip were homogenised by using \texttt{DAOMATCH} and \texttt{DAOMASTER}, and their weighted mean and standard deviation were finally adopted as star magnitude and photometric error.
The final catalog includes all the sources detected in at least two filters.%, which corresponds to about 46000 stars.

Instrumental magnitudes were calibrated 
%by using the stars in common with the catalog presented in \citet{saracino20chromo} with the only exception of the F438W magnitudes that are not available in their catalog and were calibrated 
by using the equations and zero points quoted in the dedicated instrument webpage\footnote{\url{https://www.stsci.edu/hst/instrumentation/wfc3/data-analysis/photometric-calibration}}.
Magnitudes were then corrected for the effect of differential reddening following the approach described in \citet[see also \citealt{dalessandro18b} and the Appendix for further details]{cadelano20a}.

Instrumental positions were corrected for filter-dependent geometric distortions using the prescriptions by \citet{anderson06,bellini09,bellini11} and then converted into the absolute coordinate systems by using the stars in common with \citet{saracino20chromo} as a secondary astrometric reference frame.

The left panels of Figure~\ref{fig:cmd} show the ($m_{F438W},m_{F438W}-m_{F814W}$) and 
the ($m_{F438W},m_{F343N}-m_{F438W}$) differential reddening corrected CMDs as an example.

\subsection{Proper motion analysis}
We took advantage of the large temporal baseline of $\sim15$ yr 
spanned by the observations and obtained over 5 different epochs (i.e. 2006, 2011, 2016, 2019 and 2021) to perform a relative proper motion (PM) analysis and clean the cluster CMD from field interlopers. 
%Disentangling the cluster population from interlopers mainly belonging to the Magellanic Clouds has been proven to be challenging even for the performances of the HST cameras due to the large distance of these systems and relative orbit of the clusters around the host galaxies. Indeed, only recently \citet{massari21} was able to solve the kinematic of the GC NGC419 from that of its host Small Magellanic Cloud.
To derive the cluster's relative PMs, we followed the approach described in \citet[see also \citealt{dalessandro18a,cadelano17,massari21}]{dalessandro13}. 
The procedure consists in measuring the instrumental position displacements of the stars detected in all the available epochs, once a common distortion-free reference frame is defined. 
%The reference frame adopted in this analysis is the 2006 ACS geometric distortion corrected catalogue.
As a first step, we obtained a precise measurement of the mean stellar positions in each epoch by averaging their instrumental coordinates measured in each frame of each filter. A $3\sigma$-clipping rejection was applied to maximize the accuracy of the final measurements. 
%For the WFC3 images (x,y) have been corrected for geometric distortions by applying the equations published in \citet{bellini09} and for the ACS catalogue we adopted the ACS/WFC Distortion Correction Tables (IDCTAB) provided on the dedicated page of the Space Telescope Science Institute.
We then used a six-parameter linear transformation to shift the average positions of all the stars to a master-list reference frame, which is composed by a sample of likely cluster's member stars selected according to their position in the optical CMDs of the 2006 ACS observations. For each star, the master-frame transformed positions as a function of the epoch are fit with a least-squares straight line, the slope of which represents the star's PM. The fitting procedure is iterated after data rejection and $\sigma$-clipping. After deriving the first-pass PM estimates, we repeated the entire procedure refining the reference master-list by selecting likely member stars according to their first-pass PMs. 

\begin{figure}[h] 
\centering
\includegraphics[scale=0.4]{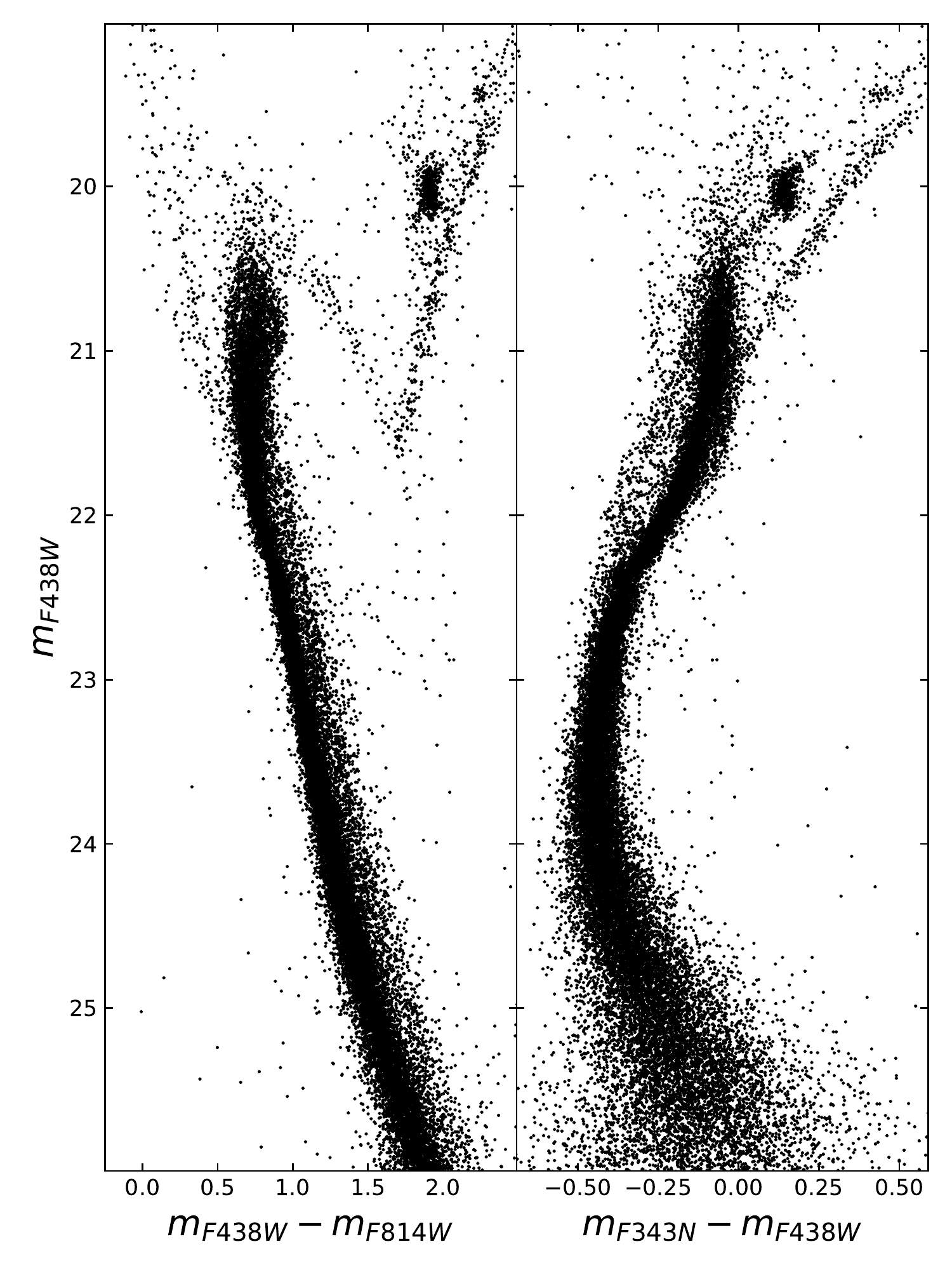}
\includegraphics[scale=0.4]{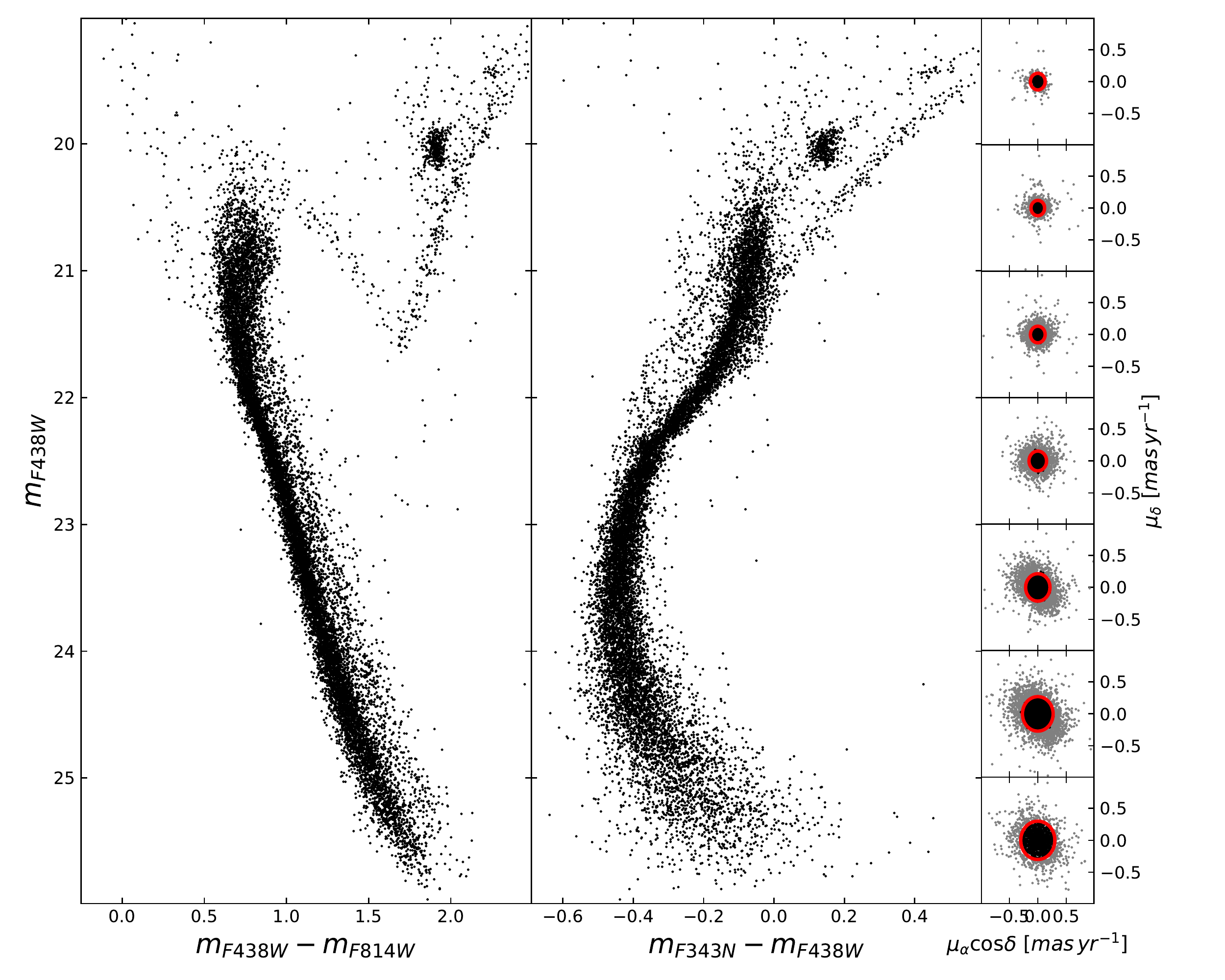}
\caption{{\it Left-hand panels:} CMD of NGC~1783 in two different combination of filters as obtained from the entire catalog of stars. Magnitudes are corrected for differential reddening. {\it Right-hand panels:} same CMD as in the left-hand panels but only for PM selected stars. The right-hand panels show the VPD in different magnitude ranges: gray points represent all the stars with a PM measurement, the red circles enclose stars  selected as bona-fide cluster's members, highlighted with black points.}
\label{fig:cmd}
\end{figure}

To obtain a catalog of stars composed of high probability cluster's members, 
we first applied quite strict astrometric quality selection criteria. Specifically, following the prescriptions by \citet{libralato19} we selected {\it i)} stars for which the reduced $\chi^2$ of the PM fit is smaller than 2 in both components, {\it ii)} stars having a PM fit based on at least three different epochs, {\it iii)} stars having a PM error smaller than $3\sigma$ (where $\sigma$ is the local standard deviation of the PM errors calculated over 0.5 large F438W magnitude bins). Then, to select bona-fide cluster's members we analysed the vector-point diagrams (VPDs) in different magnitudes bins in the range $19<m_{F438W}<26$. In each VPD, we performed a Gaussian fit to both the PM components. Stars having a PM smaller than $1\sigma$, where $\sigma$ represents the best-fit gaussian width, are marked as bona-fide cluster's members and they are shown in the VPDs on the right panels of Figure~\ref{fig:cmd}. 
%These stars represent the sample the following analysis is based on.

\section{Results}

\begin{figure}[h] 
\centering
\includegraphics[scale=0.6]{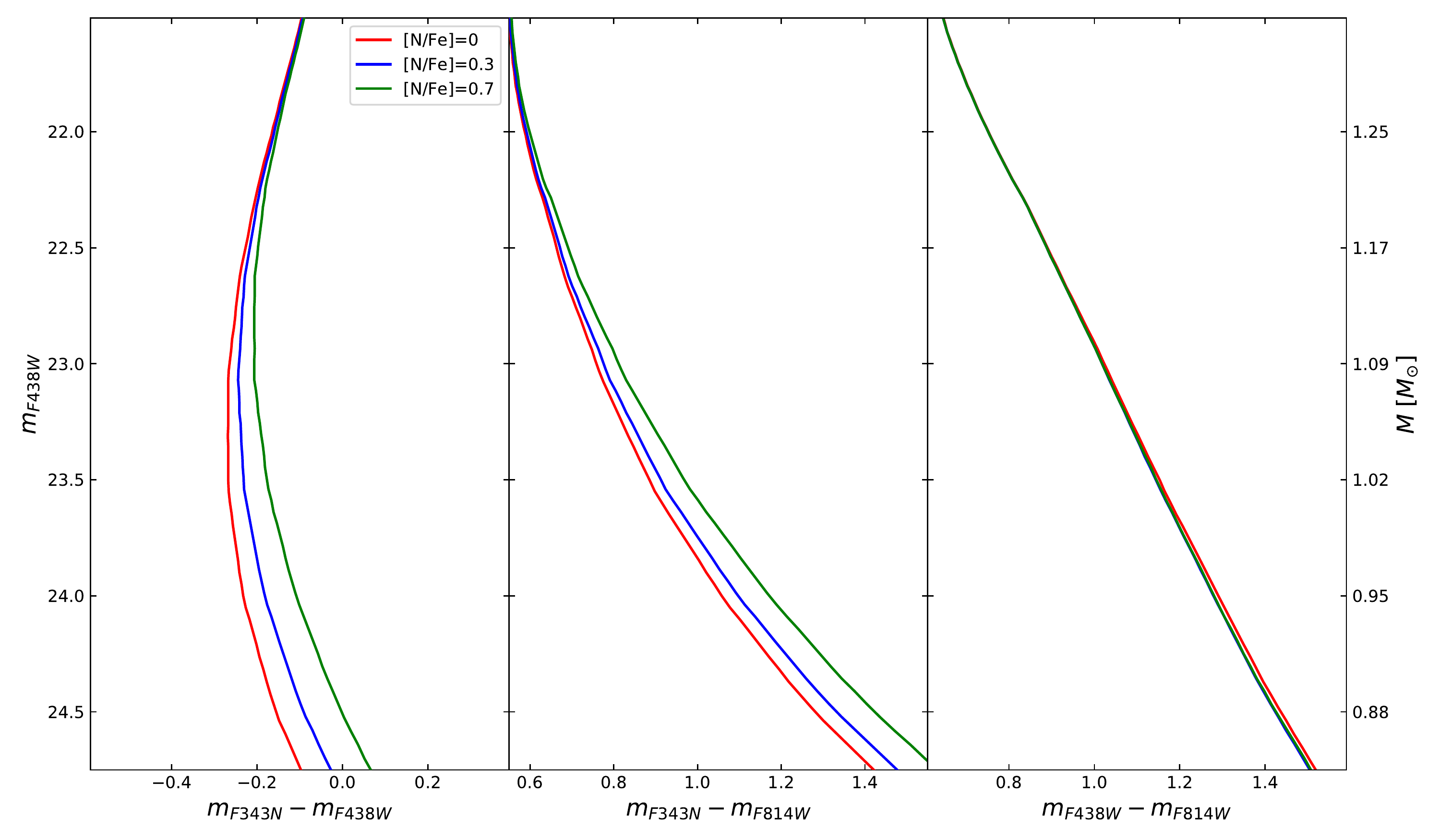}
\caption{Isochrones of a $\sim1.5$ Gyr stellar population with $[Fe/H]=-0.35$ and different N enrichment in the different filter combinations. The red, blue and green curves represent a model with a solar-scaled composition, $[N/Fe]=0.3$ and $[N/Fe]=0.7$ mixture, respectively. {The corresponding stellar masses are reported on the right-hand axis of the right-hand panel}.}
\label{fig:iso}
\end{figure}

To explore the presence of MPs along the cluster MS, we mainly exploited the F343N magnitudes, which have been shown to be quite effective in seperating MPs \citep[e.g.][]{martocchia18,cabrera-ziri20}.  In Figure~\ref{fig:iso}
we show the expected behavior of three stellar models in three example color combinations. The reference model is a BaSTI-IAC isochrone \citep{hidalgo18} of appropriate age for NGC~1783 ($t\sim1.5$ Gyr), metallicity  $[Fe/H]=-0.35$ \citep{mucciarelli08}, distance $(m-M)_0=18.45$, extinction $E(B-V)=0.02$ and scaled-solar chemical mixture. Such an isochrone is representative of the FP chemical composition. 
The other two models were obtained by using a coeval isochrone calculated for the same metallicity, but with two different choices for the metal distribution, in which the elements C, N, O follow the observed MP (anti-)correlations. Specifically, the mildly-enhanced model was obtained assuming [C/Fe]$=-0.2$, [N/Fe]$=+0.3$ and [O/Fe]$=-0.1$, while the highly-enhanced model was created assuming [C/Fe]$=-0.2$, [N/Fe]$=+0.7$ and [O/Fe]$=-0.5$. In both models the total C+N+O abundance is constant. {The calculation of the model atmospheres and fluxes has been performed as described in \citet{hidalgo18}}. Figure~\ref{fig:iso} shows that the ($m_{F343N}-m_{F438W}$) color is the most efficient combination to separate MS stars with different N abundances, in particular those with a mild N enhancement. It is also evident that the effect of N-enhancement on the ($m_{F343N}-m_{F438W}$) color becomes particularly significant for magnitude $m_{F438W}>23$. The enhancement effect decreases but is still appreciable when the F343N filter is combined with other of optical filters (see an example in the middle-panel of Figure~\ref{fig:iso}). On the contrary, as expected, the three models do not show any significant difference in the case optical filter combinations (see the example of the $m_{F438W}-m_{F814W}, m_{F438W}$ CMD on the right-hand panel of Figure~\ref{fig:iso}).  

\begin{figure}[h] 
\centering
\includegraphics[scale=0.5]{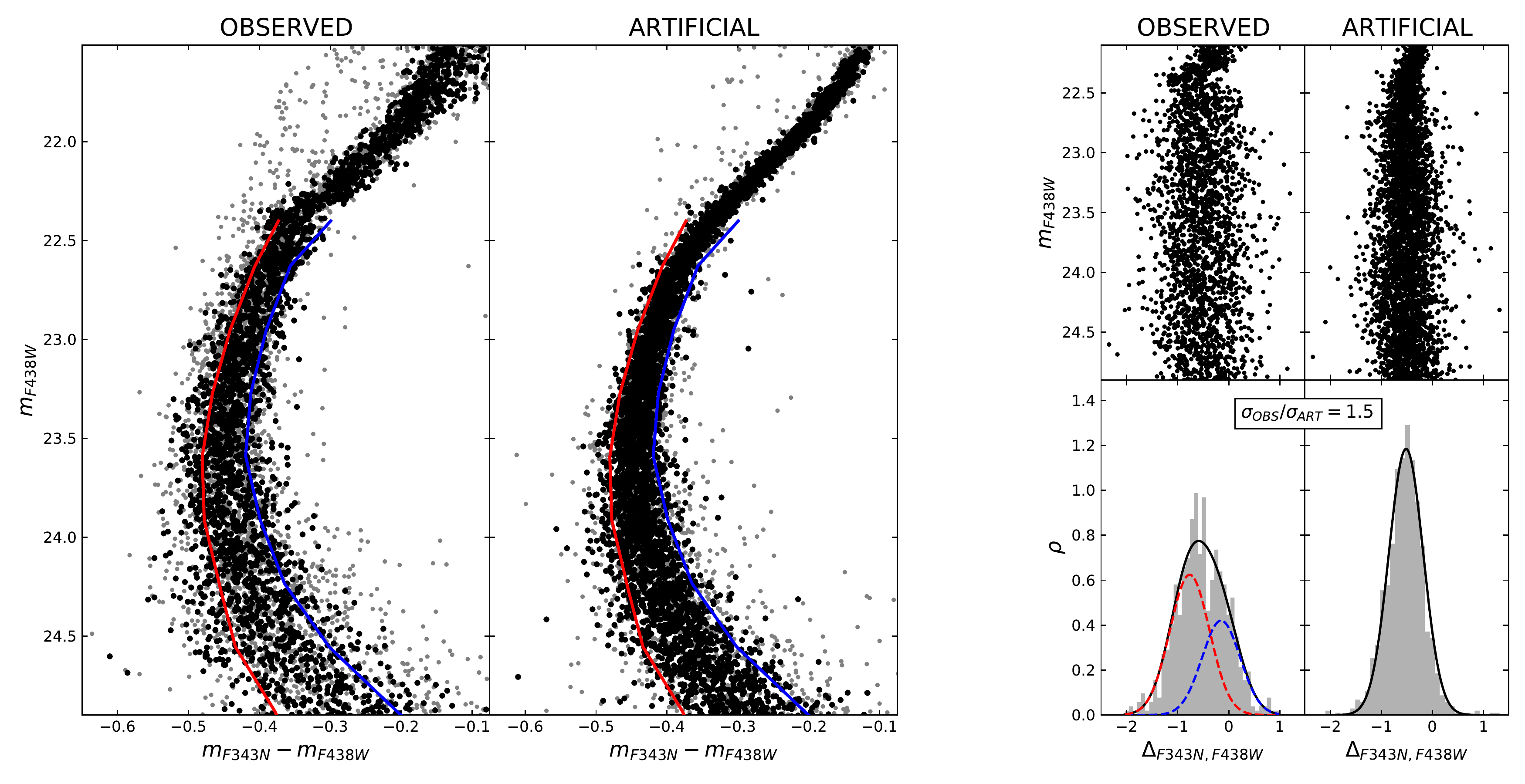}
\caption{{\it Left-hand panels}: CMD of NGC~1783 in ($m_{F343N}-m_{F438W},m_{F438W}$) filter combination of observed (left-panel) and artificial stars (right-panel). Black points are stars selected following the photometric and binary selection criteria explained in the text, while gray points are the stars that did not survive the selection. The blue and red curves are the fiducial lines adopted to verticalize the color distribution. {\it Right-hand panels:} the top panel displays the verticalized color distribution of MS stars (observed stars on the left, artificial stars on the right), while the bottom panel shows the corresponding histograms in the magnitude range $m_{F438W}=23.75-24.5$. In the case of the observed stars, the two dashed curves are the two best-fit Gaussian while the solid black curve is their sum. In the case of artificial stars, the best single Gaussian fit is shown together with the ratio between the standard deviation of the observed ($\sigma_{OBS}$) and artificial ($\sigma_{ART}$) verticalized MSs.}
\label{fig:bimod}
\end{figure}

The MP analysis was performed on stars with high photometric quality. First, we removed from the catalog stars having large photometric errors, $\chi^2$ and sharpness values. In particular, for each filter we divided the observed magnitude range in 0.5 mag large bins and removed those stars having at least one of the above quantities larger than $1\sigma$ from the local median values. 
Then we removed photometric binaries from the sample. To do this, we selected in the optical diagram ($m_{F438W}-m_{F814W},m_{F438W}$) MS stars in the magnitude range $21.5<m_{F438W}<26$. Then, we divided the sequence in 0.5 mag large magnitude bins where we evaluate the median and standard deviation of the color and removed all the $1.5\sigma$ outliers. This allows us to remove a large fraction of photometric binaries having relatively high mass ratios. 
The resulting sample is shown with black dots in Figure~\ref{fig:bimod}.

A first inspection of the CMDs in Figure~\ref{fig:cmd} confirms the presence of an extended turn-off as commonly observed in young stellar systems 
and typically interpreted as due to stellar rotation.  
Such an effect progressively fades for increasing magnitudes
and the MS reaches a minimum broadening at $m_{F438W}=22.4$ (corresponding to a mass of $\sim1.2 \Msun$). 
Interestingly, for magnitudes $m_{F438W}>22.4$ the MS width in the ($m_{F343N}-m_{F438W},m_{F438W}$) diagram abruptly starts to grow again (Figure~\ref{fig:bimod}). We note that this effect is not observed in optical CMDs, but only when the F343N band is adopted {  and therefore it is plausible to exclude this is only due to photometric errors. In addition, we can also exclude this effect is due to a residual contamination by low-mass ratio unresolved binaries, as, given the almost vertical shape of the MS in the considered magnitude range in the ($m_{F343N}-m_{F438W},m_{F438W}$) CMD, they are not expected to contribute significantly to the MS color distribution, while their effect would be more easily detectable in the optical CMDs. This points to a possible 
connection with the presence of MPs.}

To assess quantitatively whether the observed MS broadening for $m_{F438W}>22.4$ in the ($m_{F438W}-m_{F814W}$) combination can be explained in terms of photometric errors,  we compared the observations with artificial stars. 
We performed a large number of artificial star test experiments following the prescriptions in \citet[see also \citealt{dalessandro15}]{cadelano20b}. We created a list of artificial stars with a F438W input magnitude extracted from a luminosity function modelled to reproduce the observed one and extrapolated beyond the limiting magnitude. Then, to each of these artificial stars, we assigned magnitudes in all the other available filters by interpolating along appropriate mean ridge lines. These artificial stars were added to the real images by using the \texttt{DAOPHOT/ADDSTAR} software and by adopting a regular grid composed of $15\times15$ pixel cells (corresponding approximately to ten times the typical FWHM of the point spread function) in which only one artificial star for each run is allowed to lie. The photometric reduction process and the PSF models used for the artificial star experiments are the same as described in Section~\ref{sec:datared}. This process was iterated multiple times. In the end about 80000 artificial stars are simulated for the entire field of view covered by the adopted data-set. The same photometric quality selection criteria used for real stars were applied to the artificial stars.

\begin{figure}[h] 
\centering
\includegraphics[scale=0.5]{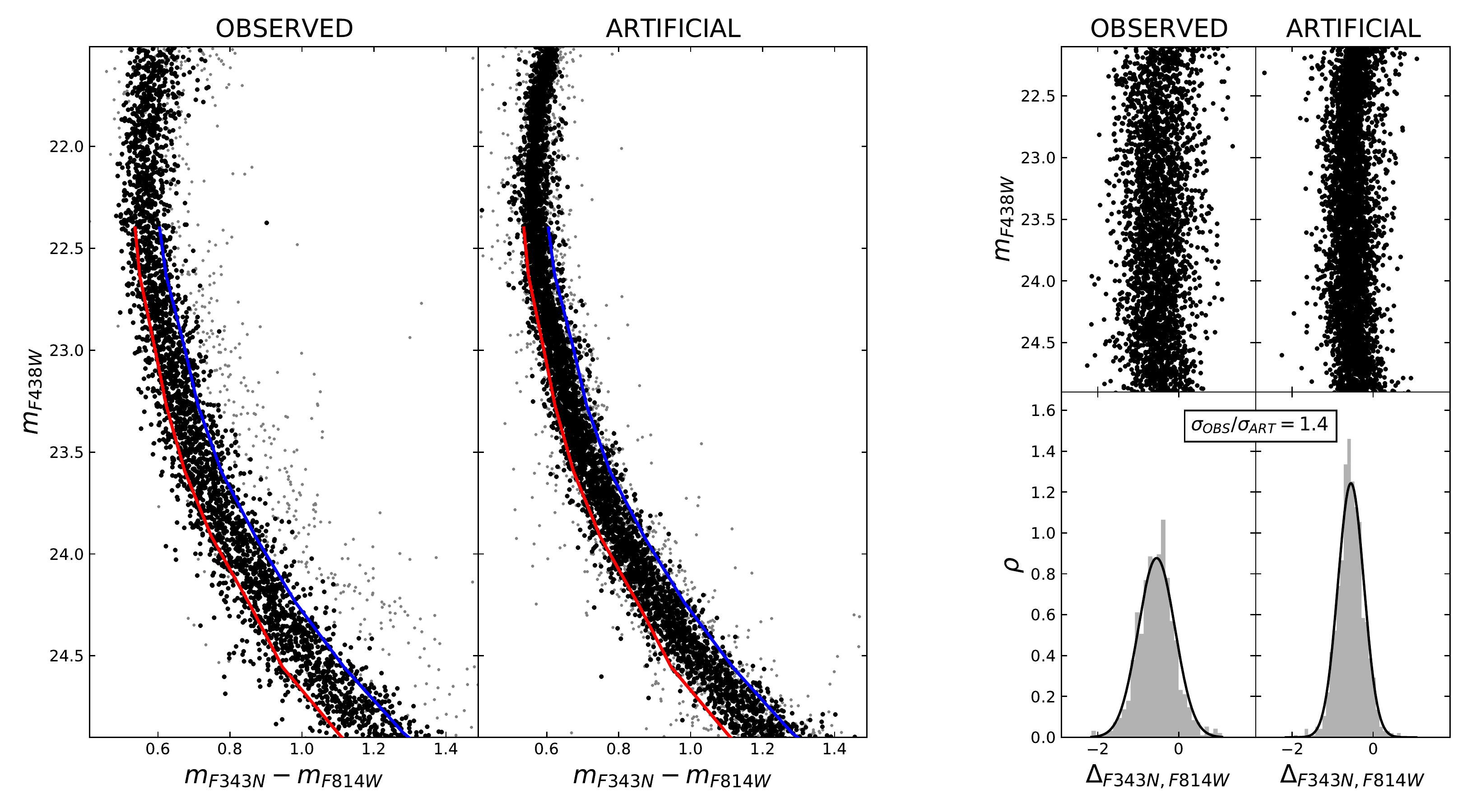}
\includegraphics[scale=0.5]{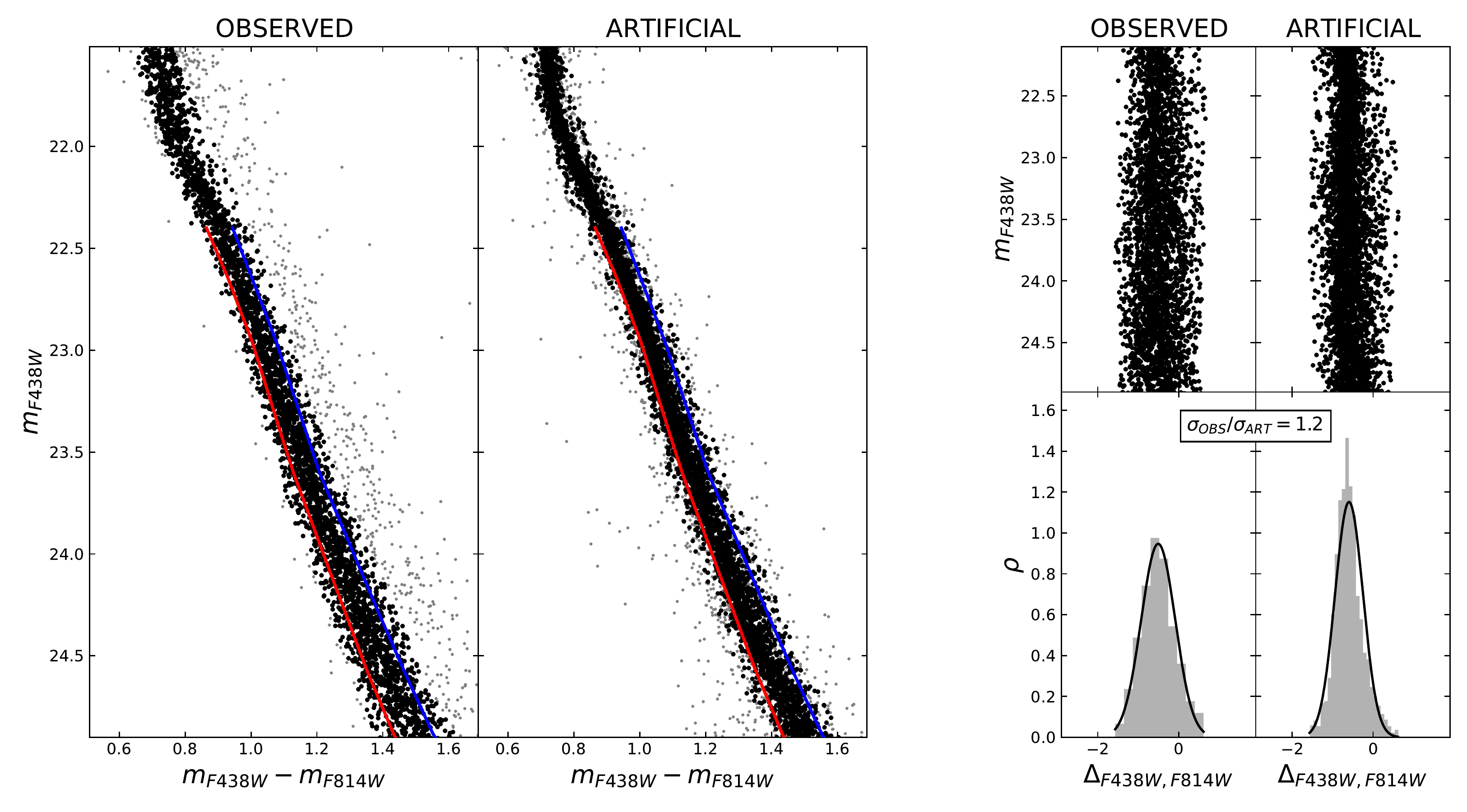}
\caption{{\it Top panels:} same as in Figure~\ref{fig:bimod} but in the case of the ($m_{F343N}-m_{F814W},m_{F438W}$) filter combination. {\it Bottom panels:} same as in the top panels but in the purely optical ($m_{F438W}-m_{F814W},m_{F438W}$) filter combination.}
\label{fig:rect}
\end{figure}

We then compared the observed MS width with that derived from artificial star CMDs.
To do this, we verticalized the distribution of MS stars with respect to two fiducial lines \citep[see][for a similar implementation of the technique]{dalessandro18b} in the magnitude range $22.5<m_{F438W}<24.5$. 
We estimated the width of the verticalized color distributions by fitting them 
with a single Gaussian function. 
%The derived standard deviation provides an estimate of the MS width. 
Results are shown in the right panels of Figure~\ref{fig:bimod}. 
Interestingly, we find that the observed MS is $\sim50\%$ larger than the artificial one. A similar difference 
is measured also when other combinations of the F343N filter with optical filters, such as the F814W filter (see top panels of Figure~\ref{fig:rect}), are considered.
On the contrary, in all the color combinations including only optical filters, such as the $m_{F438W}-m_{F814W},m_{F438W}$ CMD in the bottom panels of Figure~\ref{fig:rect}, 
the observed verticalized distributions have only a $15-20\%$ larger widths than the artificial ones. It is important to stress that such an effect is commonly observed in this kind of comparisons (see for example \citealt{dalessandro11,milone12}) and therefore cannot be considered as an evidence of a significant difference.  

%We repeat the same analysis in all the other available filter 
%combinations and we provide an example in the lower panels of Figure~\ref{fig:rect}.

%%%%%%%%%
%In general, the artificial verticalized distributions usually have a $\sim10-15\%$ smaller widths than the observed ones in all the available filter combinations not including the F343N filter.  Such trend is commonly observed in this kind of analysis (see for example \citealt{dalessandro11,milone12}. 
%On the contrary, in the case of the F343N-F438W filter combination, we find that the observed main sequence is $\sim50\%$ larger than the artificial one. We stress that such a difference is significantly larger than that observed in all the other filter combination not sensible to N variations. Interestingly, a $\sim50\%$ of difference is measured also when combinations of the F343N filter with other optical filters, such as the F814W filter. At odds with the purely optical filter combinations, these combinations are expected to be sensitive to some extent to N variations. 

{\it This quantitative analysis suggests that the significant broadening along the MS of NGC~1783, observed only when UV filters N-abundance sensitive combinations are considered, can represent the first detection of MPs in a massive stellar cluster younger than 2 Gyr}.

\begin{figure}[h] 
\centering
\includegraphics[scale=0.5]{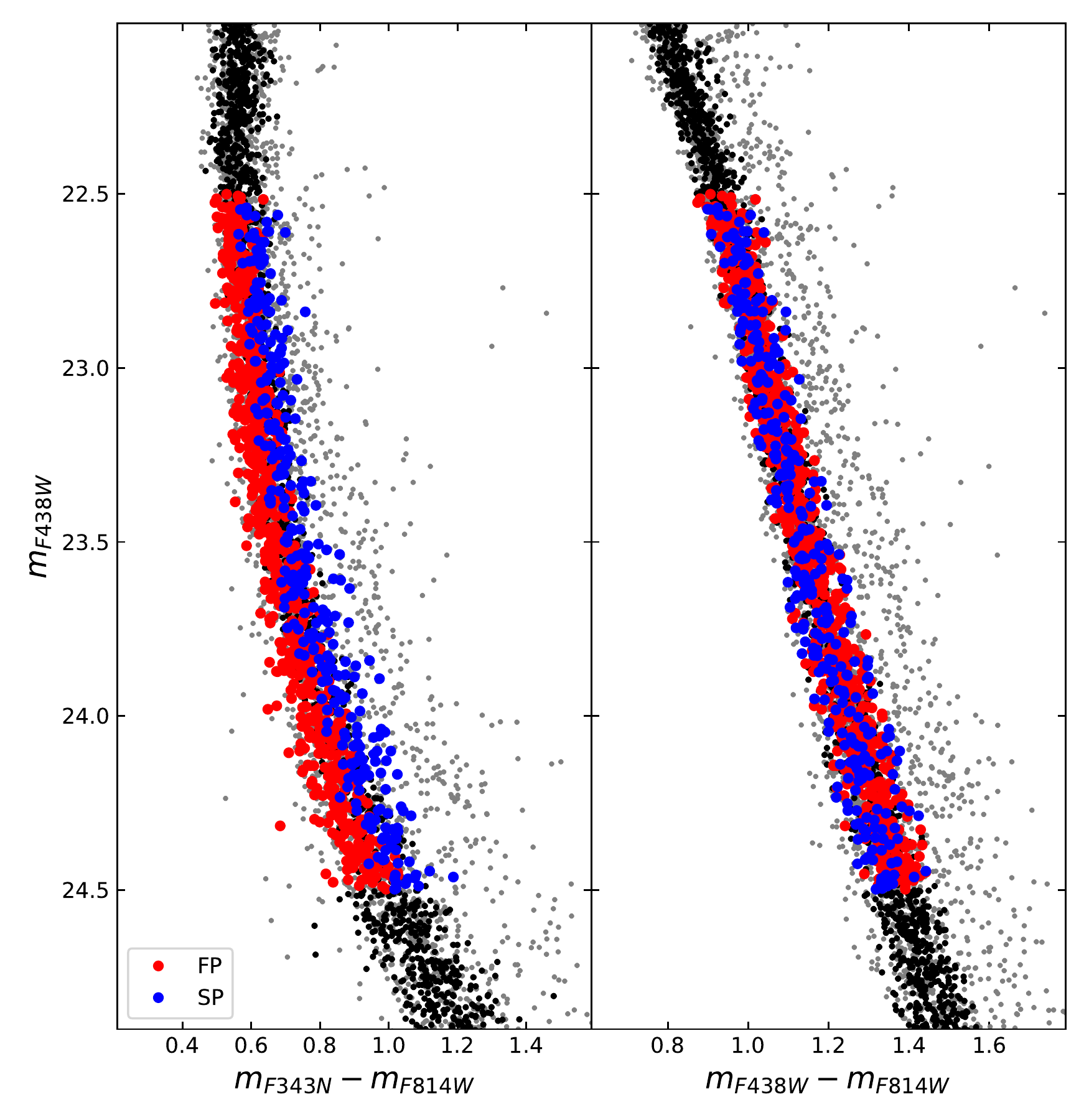}
\caption{{\it Left-hand panel:} CMD of NGC~1783 in the filter combination $(m_{F343N}-m_{F814W},m_{F438W})$. Black and gray points are the cluster's members that survived and did not survived the photometric selection criteria. Red and blue dots are FP and SP stars selected on the basis the bimodality presented in Figure~\ref{fig:bimod}. {\it Right-hand panel:} same as in the left-hand panel but in the purely optical filter combination $(m_{F438W}-m_{F814W},m_{F438W})$.}
\label{fig:bimod2}
\end{figure}

Such observational evidence is further supported by the fact that the verticalized ($m_{F438W}-m_{F814W}$) color distribution (Figure~\ref{fig:bimod}) shows hints of bimodality. Indeed, in the observed star histogram in Figure~\ref{fig:bimod} we can distinguish two distinct peaks 
with $\Delta_{F343N,F438W}\sim 1$ mag, that can be nicely fit by two Gaussian functions\footnote{We used the Gaussian Mixture Model statistics (\url{https://scikit-learn.org/stable/index.html}) to perform the two component fit.}, whose width is compatible with that expected from photometric errors (i.e. their widths are compatible with that derived from artificial stars). {  We can exclude that such bi-modal distribution can be due to low-mass ratio unresolved binaries as they are expected to uniformly populate the the MS color distribution in the considered range of magnitudes.}
Based on the expected distribution of MPs in this CMD (see Figure~\ref{fig:iso}), the red Gaussian peak corresponds to the FP stars, and includes $\sim60\%$ of the sample, while the bluer one corresponds to SP stars and includes the remaining $\sim40\%$ of objects. We note that the presence of MPs along the MS becomes apparent in the magnitude range populated by stars with mass $M\leq1 \, M_{\odot}$ (Figure~2). Finally, it is worth stressing that FP and SP stars are nicely separated in all color combinations including the F343N band, while they become indistinguishable when optical filter combinations are considered (Figure~\ref{fig:bimod2}).

\subsection{Comparison with theoretical models}
\label{sec:models}
\begin{figure}[h] 
\centering
\includegraphics[scale=0.55]{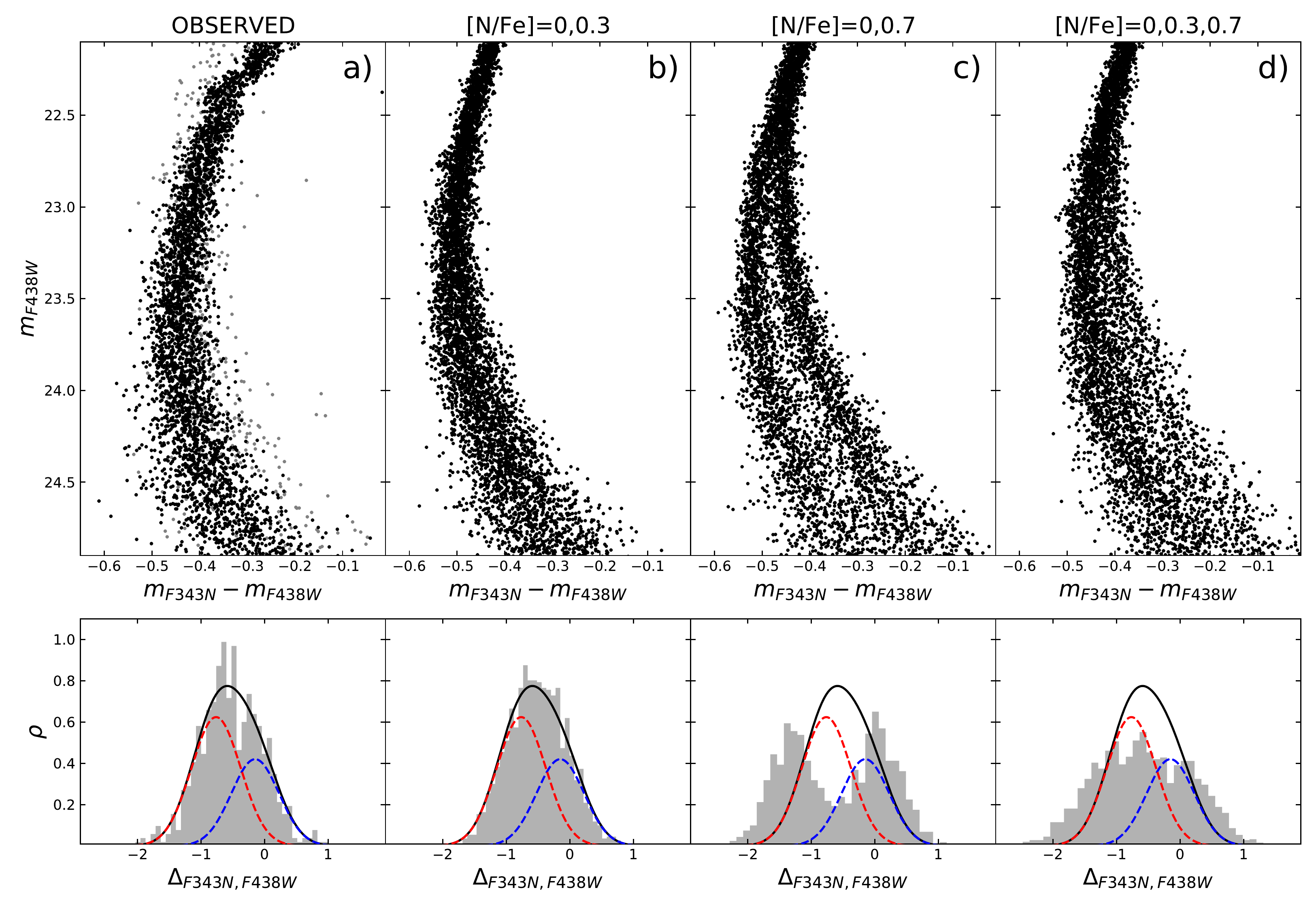}
\caption{ {\it Panel a):} observed CMD in the ($m_{F343N}-m_{F438W},m_{F438W}$) filter combination. {\it Panel b):} synthetic CMD obtained with a mixture of $[N/Fe]=0$ and $[N/Fe]=0.3$ stars. {\it Panel c):} same as in panel b) but for a mixture of $[N/Fe]=0$ and $[N/Fe]=0.7$ stars. {\it Panel d):} same as in panel b) panel but for a combination of the three available mixtures. {  {\it Bottom panels:} histograms of the verticalized distributions of MS stars in the corresponding top panels. The blue, red and black curves are the two Gaussian fit presented in Figure~\ref{fig:bimod}.}}
\label{fig:synth}
\end{figure}

To tentatively quantify the degree of N enrichment between FP and SP stars, we compared the observations with a set of synthetic CMDs mimicking a population composed of a mixture of stars having standard solar-scaled chemical composition and stars having a N enriched composition. %this, first we simulate synthetic spectra of stars with metallicity  $[Fe/H]=-0.35$ and three different N abundances: $[N/Fe]=0$, $[N/Fe]=0.3$ and $[N/Fe]=0.7$. For each chemical mixture we generate synthetic spectra in an appropriate temperature and surface gravity grid (COME???ALESSIO?). These spectra are then used to calculate bolometric corrections, which have been applied to an isochrone obtained from the BaSTI database \citep{hidalgo18} and reproducing a 1.8 Gyr old population at the cluster's metallicity. Absolute magnitudes are transformed to the observed frame by assuming an absolute distance modulus of 18.45 \citep{mucciarelli07} and $E(B-V)=0.02$. The three different isochrones are presented in the left-hand panel of Figure~\ref{fig:synth}. 
To do so, we generated three different synthetic CMDs by populating the three isochrones shown in Figure~\ref{fig:iso}. 
%one reproducing a population with solar-scaled and intermediately enriched $[N/Fe]=0.3$ stars, one with solar-scaled and highly enriched
%we created synthetic CMDs by populating the three isochrones shown in Figure~2 with artificial stars. We generate three different synthetic CMDs: one reproducing a population with solar-scaled and intermediately enriched $[N/Fe]=0.3$ stars, one with solar-scaled and highly enriched $[N/Fe]=0.7$ stars and finally one with a combination of the three mixtures. 
We divided the magnitude range $21.5<m_{F438W}<26$ in regular bins of 0.5 mag width. In each bin and for each isochrone we simulated 50 artificial stars by randomly extracting them from an uniform distribution in magnitude and from a normal distribution centered on the isochrone color and with a standard deviation equal to that measured from the artificial stars in the same magnitude bin. 
Here we assumed a flat luminosity function for the synthetic population and equally populated FP and SP. We note that results are basically unchanged if slightly different luminosity functions and population ratios are assumed.
Results are shown in Figure~\ref{fig:synth}.

The synthetic CMDs obtained by including stars with $[N/Fe]=0.7$ (Figure~\ref{fig:synth}) show either a clear split MS or a significantly larger ($>50\%$) broadening with respect to the observed CMD.
On the contrary, we find that the synthetic CMD populated by a mixture of solar-scaled and $[N/Fe]=0.3$ stars is able to nicely reproduce the observations. In fact, the resulting MS width differs by only $\sim10\%$ from the observed one {  and the histogram of the verticalized distribution nicely matches the observed color distribution (Figure~\ref{fig:bimod} and Figure~\ref{fig:synth}) - panel a)}. This suggests that NGC~1783 hosts a second population of stars moderately enriched in terms of N ($\Delta([N/Fe]\sim0.3$). 

\section{Discussion}
The observational results presented in this Letter show that we have detected for the first time the presence of MPs differing in terms of their light element abundances in a stellar cluster younger than $\sim2$ Gyr. These findings represent a potential major breakthrough in the field, as they would suggest, at odds with what found in the literature so far, that the MP phenomenon is common to all massive clusters, irrespective of their age. {Hence, if GC formation is not a specific phenomenon of stellar systems formed at high-z, then GCs at any age can be used as a proxy to study the galaxy assembly processes \citep{horta21,kruijssen15}}.

Recently, 
\citet{cabrera-ziri20} and \citet{li20} have carried out a similar study, looking for MPs on 
the MS of the massive $\sim1.5$ Gyr old cluster NGC 419 in the SMC {(see also the case of NGC~1846, \citealt{li21})}. The authors did not find any significant evidence of MPs in this cluster, however, their results might have been hampered by the quality of the available photometry (mainly due to the larger distance of the system) and therefore the cluster would deserve a follow-up analysis.

By comparing the observed CMDs with artificial stars and by following broadly the same approach used for the MS analysis, we confirm, based on more accurate photometry, previous findings about the lack of MPs along the cluster RGB. 
However, we estimate that the apparent disagreement between the results obtained for the RGB and MS is compatible with the expected mixing effects linked to the first dredge-up. 
In fact, the results presented in \citet{salaris20} show that the mixing associated to the first dredge-up can reduce the initial N differences among different sub-populations by a factor of about 2-3 at an age of $\sim1.5$ Gyr. Therefore, in the specific case of NGC~1783, an initial spread of $\Delta [N/Fe] \sim 0.3$ dex, as constrained from the MS (Section 3.1), would be erased completely on the RGB, thus mimicking an homogeneous stellar population.

The combination of these results therefore suggest that to study MPs in very young systems it is necessary to focus on their MS, thus largely changing the observing strategies adopted so far.  This calls for a dedicated study that would reappraise our understanding of the MP phenomenon over the cosmic time.
It is interesting to note in this respect, that while this work shows that there is not a sharp age limit for the onset of MPs, 
nevertheless age can indeed play a role in shaping light-element chemical abundance variations. 
If we compare the initial N spread constrained from the MS of NGC~1783 ($\Delta(N/Fe])\sim 0.3$ dex) with what found photometrically from the RGBs of intermediate-age and old clusters, after accounting for the effects of the first dredge-up, we find indications of a possible correlation between cluster age and initial N spread with older clusters requiring an initial internal $N$ variation of $\sim1$dex and the young ones a spread smaller by a factor $\sim5$ {(  see Figure~6 in \citealt{salaris20} and references therein)}.   
While it is necessary to investigate the significance of this trend further, one possibility is that it might be related to the initial cluster mass. In fact, while all clusters analyzed so far have comparable present-day masses ($M>10^5 M_{\odot}$), older clusters could have been much more massive at birth than the younger ones. 
Larger masses may allow GCs to retain more efficiently stellar ejecta, and also accrete 
pristine gas from their surroundings. However, the notion that GCs lose a significant fraction of their initial mass or are able to accrete/retain significant amounts of gas is still strongly debated (e.g., \citealt{larsen14,bastian15,cabrera15,dalessandro19}).    

{The results presented here also open the possibility of tightly constraining MP formation processes. For example, young star clusters can be used to detect the presence of age spread (not possible in the case of old clusters) which is one the major discriminator among MP formation models \citep[e.g.][]{martocchia18,martocchia19,saracino20muse}.}

A detailed characterization of the MP properties in NGC~1783 requires a detailed spectroscopic follow-up.
Given the distance of the system and faint magnitudes of the target stars, the use of integral field spectrographs and the application of the approach successfully adopted by \citet{latour19} and \citet{saracino20muse} appears to be a promising route.  

%nothing changes if different PM cuts are applied 

\begin{acknowledgments}
MC and ED acknowledge financial support from the project Light-on-Dark granted by MIUR through PRIN2017-2017K7REXT. MS acknowledges support from the STFC Consolidated Grant ST/V00087X/1.  
\end{acknowledgments}

\vspace{5mm}
\facilities{HST(ACS,WFC3)}

\software{DAOPHOT IV \citep{stetson87,stetson94}         
        }



\appendix

\section{Differential reddening}
We corrected the observed magnitudes for the effects of differential reddening following the same approach described in \citet[see also \citealt{dalessandro18b}]{cadelano20a}. Briefly, we selected a sample of RGBs in the F438W$-$F814W frame and created a mean ridge line in the magnitude range $19.5<m_{F438W}<21.5$. Then we computed the distance of each one of these selected stars from the mean ridge line along the reddening vector, defined using the extinction coefficients obtained from \citet{cardelli89} and \citet{girardi02}. 
This reference sample is used to assign a distance from the mean ridge line to all the sources in our photometric catalog, calculated as the $\sigma$-clipped median of the distance values measured for the 30 closest reference stars. Finally, the resulting values of the distances can be easily converted into variation of the color excess $\delta E(B-V)$ using an adapted version of equation~1 in \citet{cadelano20a}.
As expected due to the low average extinction and very well defined CMD sequences, we find negligible reddening variations ($\delta E(B-V) \leq 0.01$) within the surveyed field of view. Stellar magnitudes in Figure~\ref{fig:cmd} are corrected for differential reddening.


\bibliography{main}{}
\bibliographystyle{aasjournal}

\end{document}